\documentclass[prd,aps,showpacs,nofootinbib,nobibnotes]{revtex4}
\newcommand{\dd}{{\rm d}}

\begin{document}
\title{`$c$' is the speed of light, isn't it?}

 \author{George F.R. Ellis}
 \email{ellis@maths.uct.ac.za}
 \affiliation{Department of Mathematics and Applied Mathematics,
             University of Cape Town,\\
             Rondebosch 7700, Capetown, South Africa}
 \author{Jean-Philippe Uzan}
 \email{uzan@iap.fr}
 \affiliation{Institut d'Astrophysique de Paris, GR$\varepsilon$CO, FRE
             2435-CNRS, 98bis boulevard Arago, 75014 Paris, France \\
             Laboratoire de Physique Th\'eorique, CNRS-UMR 8627,
             Universit\'e Paris Sud, B\^atiment 210, F-91405 Orsay
             c\'edex, France}
\date{\today}
\begin{abstract}
Theories proposing a varying speed of light have recently
been widely promoted under the claim that they offer an
alternative way of solving the standard cosmological problems.
Recent observational hints that the fine structure constant may
have varied during over cosmological scales also has given impetus
to these models. In theoretical physics the speed of light, $c$,
is hidden in almost all equations but with different facets that
we try to distinguish. Together with a reminder on scalar-tensor
theories of gravity, this sheds some light on these proposed
varying speed of light theories.
\end{abstract}
\maketitle
\section{Introduction}

Recent observational claims, based on the measurement of distant
quasar absorption spectra, that the fine structure constant may
have been smaller in the past~\cite{webb} have restarted the
debate on the nature of the constants of nature.

On the experimental side, many new (and sharper) constraints from
a large variety of physical systems on different time scales have
been determined (see Ref.~\cite{jpu,jpu3} for recent reviews). On
the theoretical side, tests of the constancy of the constants of
Nature extend the testing of the Einstein equivalence principle to
astrophysical scales. The proof that some constants have varied
during the history of the universe will be a sign of the existence
of a new force that will most probably be composition
dependent~\cite{jpu,jpu2}. Many theoretical motivations for such
variation, mostly in the framework of higher dimensional theories
such as string theory, have been put forward~\cite{string}. Among
these models, the varying speed of light (VSL) models have been
argued to be an alternative way (compared to inflation) to solve
the standard cosmological problems~\cite{am,bm,bm2,barrow} and a
recent study on black holes suggested~\cite{bh} that a variation
of the speed of light can be discriminated from a variation of the
elementary charge (see however Ref.~\cite{nobh} for a clear
refutation of this claim).

It is however well known that only the variation of dimensionless
quantities can be determined~\cite{jpu,trialogue}, mainly because
measuring a physical quantity reduces to comparison with a physical
system that is chosen as reference. Clearly, the values of the
standard units depend on historical definitions and the numerical
values of the constants of nature are somehow dependent on these
choices. What is independent of the definition of the units are
dimensionless ratios that e.g. characterize the relative size,
strength etc. of two objects or forces (see Section~II.B of
Ref.~\cite{jpu} for a detailed discussion of metrology and
measurements and the link with the constants of nature). However,
while only variation of dimensionless constants is meaningful, one can
implement a theory in which such dimensionless constants are varying
by assuming that a dimensional constant is varying while specifying
clearly what other quantities are being kept fixed. An example is
Dirac theory~\cite{dirac} in which the gravitational constant varies
as the inverse of the cosmic time but in atomic units, so that in
particular the electron mass, $m_e$, is fixed.  It follows that this
corresponds to a theory in which the dimensionless quantity
$Gm_e^2/\hbar c$ is varying.  This theory can be considered as
resulting from atomic clocks varying relative to gravitational clocks
\cite{can}.

Before we turn to the specific case of the speed of light, let us
recall that the role and status of the fundamental constants of
physics have been widely debated. Here we shall define these
constants as all the physical parameters that are not determined
by our theory at hand. They are fundamental in the sense put
forward by Steven Weinberg~\cite{weinberg}: ``we cannot calculate
[them] with precision in terms of more fundamental constants, not
just because the computation is too complicated [...] but because
we do not know anything more fundamental". The constants that can
be included in such a list can be split (in a non-unique way) into
two sets: a set of dimensionless ratios (that can be called
fundamental parameters and are pure numbers) and a list of
dimensional constants (that can be called, following a proposal by
Okun~\cite{trialogue}, fundamental units). How many such
fundamental units are needed is still debated. To build on this
debate~(see Ref.~\cite{trialogue} for different views), let us
recall a property of the fundamental units of physics that seems
central to us: each of these constants has acted as a ``concept
synthesizer"~\cite{levy,lu}, i.e. it unified concepts that were
previously disconnected into a new concept. This for instance
happens in the case of the Planck constant and the relation
$E=\hbar\omega$, that can be interpreted not as a link between two
classical concepts (energy and pulsation, or in fact matter and
wave) but rather as creating a new concept with broader scope, of
which energy and pulsation are just two facets. The speed of light
also played such a synthesizing role by leading to the concept of
space-time, as well as (with Newton's constant) creating the link,
through the Einstein equations, between spacetime geometry and
matter (see Refs.~\cite{levy,lu} for further discussion). These
considerations, as well as facts on the number of independent
units needed in physics~\cite{jpu,trialogue}, tend to show that
three such quantities are needed. It also leads, when investigated
backward, to the concept of the cube of physical
theories~\cite{okun}.

Indeed, the numerical values of these fundamental units are
arbitrary, and can even be chosen to be unity in an adapted
natural units system, but their ``concept synthesizer" role
remains. This has the consequence that these constants relate
different concepts and thus play an intricate role in physical
laws. If one were to relax their being constants then one would
also need to relax the synthesis they underpinned. It follows that
one would need to develop a careful conceptual framework to
implement their possible variation. Our goal is to discuss, within
this understanding, a class of theories that have recently
received a lot of attention: the varying speed of light theories.

To examine the role of the speed of light, $c$, in physics, we
start as a warm-up by recalling briefly its biography in
Section~\ref{sec_history} (further details and references can be
found in Ref.~\cite{lu}) and then try, in Section~\ref{sec_faces},
to describe the various facets of the speed of light in physical
theory. In Section~\ref{sec_ST}, we recall some well-known results
concerning scalar-tensor theories of gravity in order to emphasize
the importance of writing and varying a Lagrangian in a consistent
way. In Sections~\ref{sec_vsl} and~\ref{sec_moffat} we then
discuss, in the light of the previous facts, some theoretical
frameworks in which the speed of light is supposed to vary,
finding a number of problems in their implementation. We conclude
in Section~\ref{sec_concl}.

\section{What is the speed of light?}

\subsection{A short biography of $c$}\label{sec_history}

During Antiquity, it was believed that our eyes were at the origin
of light and that its speed was infinite. One had to wait for
Galileo for this view to start changing. He was the first to try
to measure the speed of light experimentally, but his experiment
was unsuccessful. Ironically, by discovering the satellites of
Jupiter in 1610, he opened the door to the first determination of
the speed of light by the Danish astronomer Ole Roemer in 1676.
This measurement was then sharpened by James Bradley in 1728 who
utilized his discovery of the aberration of light.

During this period the status of the speed of light was somehow
not different from that of the speed of sound: it was simply a
property of light. Huygens proposed a description in terms of
waves, contrary to Newton's corpuscular description. This wave
description was backed up experimentally by Foucault in 1850 who
checked that the speed of light was smaller in a refractive medium
than in vacuum. Quite naturally, this lead almost all physicists
to believe that light requires a medium to propagate in, which was
named {\it ether}. The speed of light was thus a property of the
ether itself, in the same way as the speed of sound can be
computed in terms of the temperature, pressure, and properties of
the gas it propagates in. Clearly, it was not thought to be
fundamental.

In 1855, Kirchhoff realized that $(\varepsilon_0\mu_0)^{-1/2}$ has
the dimension of a speed, where $\varepsilon_0$ and $\mu_0$ are
two constants entering the laws of electricity and magnetism.
Weber and Kohlrausch measured this constant in 1856, using only
electrostatic and magnetostatic experiments. Within the
experimental accuracy, it agreed with the speed of light. This
remained a coincidence until Maxwell formulated his theory of
electromagnetism in 1865 where he concluded that ``light is an
electromagnetic disturbance propagated through the field according
to electromagnetic laws''. At that stage, the status of $c$
increased tremendously since it became a characteristics of all
electromagnetic phenomena. Note that it is not only related to a
velocity of propagation since it can be measured by some
electrostatic and magnetostatic experiments.

The next mutation of $c$ arose from the incompatibility of
Maxwell's equations with Galilean invariance: either one sticks to
Galilean invariance and Maxwell's equations only hold in a
preferred frame, so that measurements of the velocity of light
should allow one to determine this preferred frame, or one has to
improve on Galilean invariance. Michelson-Morley experiments,
under Einstein's interpretation, implied one should follow the
second road: Galilean invariance was replaced by Lorentz
invariance, and $c$ triggered the (special) relativity revolution.
It was becoming the link between space and time and so entered
most of the laws of physics, mainly because it enters the notion
of causality. Thus for example it became later apparent that it
also is the speed of propagation of gravitational waves, and
indeed that of any massless particle.

The last, but not least, change happened in 1983 and was driven by
experimental needs. The accuracy of the experimental determination
of $c$ was limited by the accuracy and reproducibility of the
realization of the meter. The BIPM thus decided to redefine this
unit in terms of a value of $c$ that was fixed by law. This
mirrored J.L. Synge's emphasis~\cite{synge} that the theoretically
best way to measure distance on macro-scales is by radar or
equivalent methods dependent on the speed of light, so that the
best units for distance were light-seconds or light-years; and
indeed that realization was then turned into technological reality
through surveying instruments such as the Tellurometer and more
recent developments such as the GPS system widely used for
navigation.

We see from this summary that the speed of light moved from a
simple property of light to the role of a fundamental constant
that enters many laws of physics that are {\it a priori}
disconnected from the notion of light itself.

\subsection{One constant with many facets}\label{sec_faces}

It follows that the speed of light is a complex quantity that
turns out to have different origins that lead to coincident
values. It may be of some use to try to distinguish its different
facets. As the previous section tends to show, it seems clear that
$c$ is not only the speed of light and more interestingly is
probably not even always the speed of light.

\subsubsection{$c_{_{\rm EM}}$: the electromagnetism constant}

To start with, let us remember the Maxwell
equations~\cite{jackson} in MKSA units
\begin{equation}
 \nabla.{\bf D}=\rho,\quad
 \nabla.{\bf H}=0,\quad
 \nabla\times{\bf H}={\bf J}+\partial_t{\bf D},\quad
 \nabla\times{\bf E}=-\partial_t{\bf B},
\end{equation}
where $\rho$ is the density of free charges and ${\bf J}$ the
current density. The displacement, ${\bf D}$, is related to the
electric field, ${\bf E}$, and the polarization, ${\bf P}$, by
${\bf D}=\varepsilon_o{\bf E}+{\bf P}$ while the magnetic field,
${\bf H}$, is related to the magnetic induction (magnetic flux
density), ${\bf B}$, and the magnetization, ${\bf M}$, by ${\bf
B}=\mu_0({\bf H}+{\bf M)}$, $\varepsilon_0$ and $\mu_0$ being
respectively the permittivity and permeability of the vacuum.
Introduce the potential by the standard definition ${\bf
E}=-\nabla \phi-\partial_t{\bf A}$ and ${\bf B}=\nabla\times {\bf
A}$ (which is unique up to a gauge transformation). In vacuum, it
is an easy exercise to show that the wave equation for an
electromagnetic wave is
\begin{equation}
 \left(\partial_t^2-c^2_{_{\rm EM}}\Delta\right)(\phi,{\bf A}) = 0
\end{equation}
where $\Delta$ is the Laplacian and with
\begin{equation}
c^2_{_{\rm EM}}\equiv\frac{1}{\varepsilon_0\mu_0}.
\end{equation}
Here $c_{_{\rm EM}}$ appears as the velocity of any
electromagnetic wave, and thus of light, in vacuum.

Calling $c_{_{\rm EM}}$ the speed of light is somewhat too
restrictive even at that level since it is characteristic of any
electromagnetic phenomena. It should be referred to as the {\it
electromagnetism constant}.

Formally, by setting $x^0=c_{_{\rm EM}}t$, the Maxwell equations
can be recast in the form
\begin{equation}
 \partial_\mu F^{\mu\nu}=j^\nu\quad\hbox{with}\quad
 A^\mu=(\phi,{\bf A}/c_{_{\rm EM}}),\quad
 j^\mu=(\rho,{\bf J}/c_{_{\rm EM}}),\quad
 F_{\mu\nu}=\partial_\mu A_\nu-\partial_\nu A_\mu,
\end{equation}
with $\mu,\nu=0..3$. Imposing the Lorentz gauge ($\partial_\mu
A^\mu=0$) we recover the propagation equation for $A^\mu$. Indeed,
here the wave equation constant is embodied in the notation since
$x^0$ is defined in terms of $c_{_{\rm EM}}$. Of course in a
relativistic framework, one would get the same equations since
Maxwell electrodynamics is Lorentz invariant by construction.

\subsubsection{$c_{_{\rm ST}}$: the spacetime constant}

The next role of the speed of light that we alluded to, is the
synthesizer between space and time. This constant enters the
Lorentz transformation and the spacetime description of special
relativity. There are many ways to derive the Lorentz
transformations (see e.g. Ref.~\cite{book} for an extensive
discussion).

Following the proposal of Ref.~\cite{levy2}, let us call the
constant that appears in the Lorentz transformations the spacetime
structure constant, $c_{_{\rm ST}}$ so that the Minkowski line
element takes the form
\begin{eqnarray}
 \dd s^2&=&-(\dd x^0)^2+(\dd x^1)^2+(\dd x^2)^2+(\dd x^3)^2\\
        &=&-(c_{_{\rm ST}}\dd t)^2+(\dd x^1)^2+(\dd x^2)^2+(\dd x^3)^2.
\label{metric}
\end{eqnarray}
It can be shown that this constant can be defined
(completely independently of electromagnetism) as the universal
invariant limit speed~\cite{book} so that speeds are not additive,
even though they may be approximately so at low speeds. The
conditions that the composition of speed (denoted $\oplus$)
satisfies are\\ (i) that it has an identity element, $O$ (i.e.,
$O\oplus u=u\oplus O=u$ for all $u$),\\ (ii) that there is a
universal element, in our case $c_{_{\rm ST}}$, such that
$c_{_{\rm ST}}\oplus u=u\oplus c_{_{\rm ST}}=u$ for all $u$,\\
(iii) the associative rule $u\oplus(v\oplus w)=(u\oplus v)\oplus
w$, \\ (iv) that the differentials $\dd(u\oplus v)/\dd u$ and
$\dd(u\oplus v)/\dd v$ should exist and be continuous in $u$ and
$v$, and\\ (v) that $\dd(u\oplus v)/\dd u>0$ and $\dd(u\oplus
v)/\dd v>0$ provided $u,v\not=0$\\ These lead to the standard
velocity combination rules of Special Relativity with $c_{_{\rm
ST}}$ as the limiting speed; see Ref.~\cite{book} for details and
other ways to derive the Lorentz transformations (interestingly
most of these constructions are based on axioms that are universal
principles, and are not determined by the properties of any
specific interaction of nature). The limiting speed is then given
by the metric (\ref{metric}) through the equation $\dd s^2=0$,
corresponding to motion at the speed $c_{_{\rm ST}}$.

Finally let us stress that the special relativity conversion
factor between energy and mass is given by $c_{_{\rm ST}}^2$ since
it arises from the study of the dynamics of a point particle (see
e.g. Ref.~\cite{book}). Thus the correct equation unifying the
concepts of energy and mass is $E = m\, c_{_{\rm ST}}^2$.

Let us now come to the case of the curved spacetime of general
relativity and the corresponding measurement of distances, and
consider for that purpose a space time with line element $\dd
s^2=g_{\mu\nu}\dd x^\mu\dd x^\nu$. Any observer can define their
proper time by the relation
\begin{equation}
 \dd s^2=-c_{_{\rm ST}}^2\dd\tau^2,
\end{equation}
which is the most natural choice since this implies the limiting speed
given by $\dd s^2=0$ is the same for all observers independently of
their position in space and time and of their motion; it also gives
the standard time dilation effect as experimentally measured for
example in the life times of cosmic ray decay products.  Clearly, we
have that
\begin{equation}
 -c_{_{\rm ST}}^2\dd\tau^2=g_{00}(\dd x^0)^2
\end{equation}
for any observer at rest relative to the coordinate system
$(x^\alpha)$. In the standard case (where $c_{_{\rm EM}}=c_{_{\rm
ST}}$), we can define the spatial distance, $\dd\ell$, between two
points with coordinates $x^i$ and $x^i+\dd x^i$ as the radar
distance given by $c_{_{\rm ST}}/2$ times the proper time measured
by an observer located at $x^i$ for a signal to go out and come
back between $x^i$ and $x^i+\dd x^i$. It will follow that (see
e.g. Ref.~\cite{landau})
\begin{equation}
 \dd\ell^2=\gamma_{ij}\dd x^i\dd x^j\qquad\hbox{with}\quad
 \gamma_{ij}= g_{ij}-\frac{g_{0i}g_{0j}}{g_{00}},
\end{equation}
with $i,j=1..3$. Thus the determination of distance requires both the
measurement of time and the use of a signal exchanged between the
distant points. In the standard case, there is no ambiguity because
one can use light that propagates at the universal speed ($c_{_{\rm
ST}}=c_{_{\rm EM}}$).  In a case where these velocities do not
coincide the determination of distance will become more involved
because light will follow a timelike (and not a null) geodesic of the
metric. Furthermore, in any situation where this standard
matter-geometry coupling does not hold, we need to be told what
geometric meaning, if any, the metric tensor has, and by what
alternative method times and spatial distances are to be determined.

\subsubsection{Agreement of $c_{_{\rm EM}}$ and $c_{_{\rm ST}}$}
The historical path was from electrodynamics to the demonstration
that the speed of light was constant (Michelson-Morley
experiments) to the Lorentz transformation and the group structure
of spacetime. Then it was realized from the study of relativistic
dynamics that any particle with vanishing mass will propagate with
the speed of light. But clearly, the speed of light $c_{_{\rm
EM}}$ agrees with the universal speed, $c_{_{\rm ST}}$, only to
within the experimental precision of Michelson-Morley type
experiments (or put differently, the photon has zero mass only
within some accuracy) and the causal cone need not coincide with
the light cone. If one were to prove experimentally that the
photon is massive then the standard derivation of relativity from
electromagnetism would have to be abandoned.

It is thus clear that one can keep the basic principles of a
metric theory of gravity and obtain a variable speed of light
($c_{_{\rm EM}}$) by modifying Maxwell equations~\cite{modmax}.
Many ways can be followed and many terms can be added to the
electromagnetic Lagrangian. In the Proca theory, one modifies
Maxwell equation to get a massive spin-1 state
\begin{equation}
 \partial_\mu F^{\mu\nu}+m^2 A^\nu=0.
\end{equation}
It follows that the Lorentz gauge condition is automatically
satisfied if the photon is massive but we have lost the freedom of
gauge transformation. Teyssandier~\cite{teys} considered terms
that do not violate gauge invariance and couple to the curvature
as
\begin{equation}
{\cal L}_{_{\rm EM}}=\frac{1}{4}(1+\xi R)F_{\mu\nu}F^{\mu\nu}
 +\frac{1}{2}\eta R_{\mu\nu}F^{\mu\rho}F^\nu_\rho
 +\frac{1}{4}\zeta R_{\mu\nu\rho\sigma}F^{\mu\nu}F^{\rho\sigma},
\end{equation}
where $R$ is the Ricci scalar. From this he concluded that
\begin{equation}
\frac{c_{_{\rm EM}}}{c_{_{\rm
ST}}}=1+\frac{(\eta+\zeta)(4U-R)}{3-(3\xi+\eta)R-2(\eta+\zeta)U}
\end{equation}
when the Ricci tensor takes the form $3R_{\mu\nu}=(4U-R)u_\mu
u_\nu+(U-R)g_{\mu\nu}$, $u^\mu$ being a unit timelike vector
field. Let us also emphasize that in quantum electrodynamics, the
photon gets an effective mass due to vacuum
polarization~\cite{vacpol} usually described by adding the
Euler-Heisenberg Lagrangian~\cite{he} to the standard
electromagnetic Lagrangian.

As we see, on the one hand, if the universal speed is fixed by
electromagnetism it is difficult to understand why other
interactions are locally Lorentz invariant, and on the other hand,
the invariance of the laws of physics under the Poincar\'e group
allows the existence of massless particles but does not imply that
every gauge boson must be massless. It is therefore important
conceptually to carefully distinguish the speed of light $c_{_{\rm
EM}}$ from the universal speed $c_{_{\rm ST}}$ that dictates the
properties of spacetime.

\subsubsection{$c_{_{\rm GW}}$: the speed of gravitational waves in
vacuum}

As long as we are in a vacuum, the Einstein equations derived from
the action
\begin{equation}
 S=\int R\sqrt{-g}\dd^4{\bf x}
\end{equation}
can be linearized around Minkowski spacetime. It can be shown that
the only degrees of freedom are massless states of spin 2 that are
gravitational waves. With the line element
\begin{equation}
 \dd s^2=-(c_{_{\rm ST}}\dd t)^2+(\eta_{ij}+h_{ij})\dd x^i\dd
 x^j,\qquad h_{ij}\eta^{ij}=0,\quad \partial_i h^{ij}=0,
\end{equation}
the linearised Einstein vacuum field equations reduce to the
propagation equation
\begin{equation}
 \left(\partial_t^2-c^2_{_{\rm ST}}\Delta\right)h_{ij} = 0
\end{equation}
so that the speed of propagation of these gravitational waves is
given by the universal speed of our spacetime, $c_{_{\rm
GW}}=c_{_{\rm ST}}$. As long as we assume that General Relativity
is valid this conclusion cannot be avoided but, were we able to
formulate a theory with light massive gravitons in their spectra
(see e.g. Ref.~\cite{gravmass} for some attempts in the braneworld
context) then the speed of propagation of gravity might be
different from the universal speed and would lead to modification
of the Newton law of gravitation on astrophysical
scales~\cite{modgrav}. Let us emphasize that there are dangerous
issues associated with the presence of extra-polarization states
of massive gravitons~\cite{discontinuite}. Among other
problems~\cite{discontinuite}, these massive gravitons have a
scalar polarization state whose coupling to matter does not depend
on the mass of the graviton and that is, at the linear level,
analogous to a Jordan-Brans-Dicke coupling with $\omega=0$. That
coupling will modify the standard relation of interaction between
matter and light by a factor 3/4 so that one will expect a 25\%
discrepancy in the tests of gravitation in the Solar System which
are actually at a level of $10^{-2}\%$~\cite{will}.

According to the previous discussion, it seems difficult to
consider a speed of gravity that differs from $c_{_{\rm ST}}$,
nevertheless, it may be good to keep an open mind and not forget
that these two speeds may differ.

\subsubsection{$c_{_{\rm E}}$: the (Einstein) spacetime-matter constant}

Let us now consider gravity coupled to matter. The Einstein field
equations
\begin{equation}
 G_{\mu\nu}=\frac{\alpha}{2} T_{\mu\nu} \label{efe}
\end{equation}
involve a coupling constant $\alpha/2=8\pi G/c^4$ between the
Einstein tensor $G_{\mu\nu}$ and the stress-energy tensor
$T_{\mu\nu}$, where $c$ has the dimensions of speed; let us call
it $c_{_{\rm E}}$. To interpret this constant, one needs to consider
the weak field limit of the field equations (\ref{efe}) in which
the spacetime metric is given by
\begin{equation}
 \dd s^2=-(1+h_{00})(c_{_{\rm ST}}\dd t)^2+\eta_{ij}\dd x^i\dd x^j.
\end{equation}
The geodesic equation of a massive particle ($u^\mu\nabla_\mu
u^\nu=0$ with $u^\mu\equiv\dd x^\mu/\dd s=(1,{\bf v}/c_{_{\rm
ST}}))$ then reduces to
\begin{equation}
\partial_t v^j=-c_{_{\rm ST}}^2\eta^{ij}\partial_i h_{00}/2.
\end{equation}
To get the Newtonian law, one needs to identify the metric
perturbation $h_{00}$ with the Newtonian gravitational potential,
$\phi$, giving
\begin{equation}
h_{00}=2\phi/c_{_{\rm ST}}^2.
\end{equation}
The second step is to compare the Einstein equation to the Poisson
equation. Since $R_{00} \sim
\partial_i \Gamma^i_{00} \sim-g^{ij}\partial_{ij}g_{00}$ we get
that
\begin{equation}
2\Delta\phi/c_{_{\rm
ST}}^2=\alpha\left(T_{00}-\frac{1}{2}g_{00}T\right)/2.
\end{equation}
For a fluid, $T_{00}=\rho c_{_{\rm ST}}^2$ so that we need
\begin{equation}
\alpha=16\pi G/c_{_{\rm ST}}^4
\end{equation}
to get the correct Newtonian limit for gravity. In the context of
general relativity, it is thus clear that $c_{_{\rm E}}=c_{_{\rm
ST}}$. It is however important to distinguish these concepts, as
will become clearer when discussing the varying speed of light
models.

\subsubsection{Conclusion}

If one wants to formulate a theory in which the speed of light is
varying, the first step is to specify unambiguously which of the
speeds identified here is varying and then to propose a
theoretical formulation (i.e. a Lagrangian) to achieve this task.
There is no reason why after relaxing the property of constancy of
the speed of light, the different facets of $c$ described in this
section will still coincide. Besides this, it is important to
clearly state which are the quantities that are kept fixed when
one or other aspect of $c$ is assumed to vary (see e.g. the
discussion in Ref.~\cite{okun}). In the following, we will discuss
some attempts at varying-$c$ theories.

\section{Interlude on scalar tensor theory of gravity}\label{sec_ST}

Before we proceed further with the speed of light, let us make
some general comments about scalar-tensor theories of gravity.
This section follows the notation and the main results about such
theories detailed in Ref.~\cite{st2}, see also Ref.~\cite{ru}. In
such theories, one considers a Lagrangian of the form~\cite{st}
\begin{equation}\label{actST}
S={\frac{1}{16\pi G}}\int\dd^{4}x\sqrt{-g}\left[F(\psi)
R-Z(\psi)~g^{\mu\nu}\partial_{\mu }\psi\partial_{\nu }\psi
-2U(\psi)\right] +S_{m}[\phi_{m};g_{\mu \nu }]
\end{equation}
where $G$ is the bare gravitational constant, which differs from
the measured one, and $R$ is the Ricci scalar of the metric
$g_{\mu\nu}$. The function $F$ is dimensionless and needs to be
positive for the graviton to carry positive energy. The dynamics
of $\psi$ depends on the functions $F$, $Z$ and $U$ but note that
$Z$ can always be set equal to 1 by a redefinition of the field
$\psi$ so that there remain only two arbitrary functions. The
matter action, $S_{m}[\phi_{m};g_{\mu\nu}]$, depends only on the
matter fields and the metric. Such a form implies that the weak
equivalence principle holds. In models describing varying
constants, this may not be the case anymore (see below).

The variation of this action gives
\begin{eqnarray}
F(\psi)G_{\mu \nu }&=& 8\pi GT_{\mu \nu
}+Z(\psi)\left(\partial_{\mu }\psi
\partial_{\nu }\psi -{\frac{1}{2}}g_{\mu \nu }(\partial_{\alpha}\psi
\partial^{\alpha}\psi)\right)\nonumber\\
&&+\nabla_{\mu}\partial_{\nu}F(\psi)-g_{\mu\nu}\nabla_\alpha\nabla^
\alpha F(\psi)-g_{\mu \nu }U(\psi) \\
2Z(\psi)\nabla_\alpha\nabla^\alpha \psi  &=&-\frac{\dd
F}{\dd\psi} \,R- \frac{\dd Z}{\dd\psi } \,(\partial
_{\alpha }\psi)^{2}+2\frac{\dd U}{\dd\psi}\ , \\
\nabla _{\mu }T_{\nu}^{\mu } &=&0\ ,
\end{eqnarray}
where the matter stress-energy tensor is defined as
$T^{\mu\nu}\sqrt{-g}=2\delta S_m/\delta g_{\mu\nu}$. All the
previous equations are written in the Jordan frame. In the action
(\ref{actST}), matter is universally coupled to the metric tensor
$g_{\mu\nu}$ so that this metric is in fact the one that defines
lengths and times as measured by laboratory rods and clocks, with
associated speed of light $c_{_{\rm ST}}$ (which we have here set
equal to unity by appropriate choice of units). All experimental
and observational data have their standard interpretation in this
frame.

Let us remember that there exists another interesting frame, the
Einstein frame, defined by performing the conformal transformation
\begin{equation}
\widetilde{g}_{\mu\nu}=F(\psi){g}_{\mu\nu};\quad
\left(\frac{\dd\varphi}{\dd\psi}\right)^2=\frac{3}{4}\left(\frac{\dd\ln
F} {\dd\psi}\right)^2,\quad A(\varphi)=F^{-1/2},\quad 2\widetilde
V(\varphi)=U/F^2
\end{equation}
so that the action (\ref{actST}) takes the form
\begin{equation}\label{actST2} S={\frac{1}{16\pi
G}}\int\dd^{4}x\sqrt{-\widetilde g} \left[\widetilde R-2\widetilde
g^{\mu\nu}\partial_{\mu }\varphi\partial_{\nu } \varphi
-4\widetilde V(\varphi)\right]
+S_{m}[\phi_{m};A^2(\varphi)\widetilde g_{\mu \nu }].
\end{equation}
With this form, the action looks like the action of general
relativity but the matter fields are now explicitly coupled to the
metric, so that for instance $\widetilde
T_{\mu\nu}=A^2(\varphi)T_{\mu\nu}$. The main difference between
the two frames arises from the fact that the spin-2 degrees of
freedom are perturbations of $\widetilde g^{\mu\nu}$ and that
$\varphi$ is a spin-0 scalar. The perturbation of $g^{\mu\nu}$
actually mix the spin-2 and spin-0 excitations.

An easy application is to derive the cosmological equations:
\begin{eqnarray}
3F\left(H^2+\frac{K}{a^2}\right)&=&8\pi
G\rho+\frac{1}{2}Z\dot\psi^2-3H\dot F+U,\\
-2F\left(\dot H -\frac{K}{a^2}\right)&=&8\pi G(\rho+P)
+Z\dot\psi^2+\ddot F- H\dot F,\\
\dot\rho+3H(\rho+3P)&=&0,\\
Z\left(\ddot\psi+3H\dot\psi\right)&=&3F'\left(\dot
H+2H^2+\frac{K}{a^2}\right)-Z'\frac{\dot\psi^2}{2}-U'
\end{eqnarray}
where $F'=\dd F/\dd\psi$ and $a(t)$ is the scale factor of the
Robertson-Walker metric assumed, $K=\pm1,0$ is the curvature
index, a dot refers to a derivative with respect to the cosmic
time $t$, and $H\equiv\dot a/a$, while $\rho$ is the matter
density and $P$ its pressure.

In the preceding framework, the universality of free-fall is not
violated and all constants but the gravitational constant are
constant. It can be further generalized by allowing different
couplings in $S_m$, e.g.
\begin{equation}
S_m=\int \dd^4x\sqrt{-g}\left(\frac{B}{4}(\psi)F^2+\ldots\right).
\end{equation}
which will described a theory in which both the gravitational
constant and fine structure constant are varying. In that case the
field couples differently to different particles so that one
expects a violation of the universality of free fall (see also
Ref.~\cite{cap} for a very general scalar-tensor theory in which the
speed of electromagnetic waves is computed).

\section{Single-metric formulation of the VSL models}\label{sec_vsl}

Single-metric varying speed of light models have been formulated in
different ways. Moffat proposed the idea in 1992~\cite{mof1} as a
solution to the same cosmological puzzles as inflation, and somewhat
different versions have more recently been widely popularized under
the claim that they were able to solve these
problems~\cite{am,bm,bm2}.  Here we take a fresh look at the latter
papers, taking into account what we learned from the former sections.

Briefly, the claim in~\cite{am,bm,bm2}  to solving the standard cosmological 
problems was built on the assumption that the Friedmann equations remain valid 
even when $\dot c\not=0$, so that
\begin{eqnarray}
\left(\frac{\dot{a}}{a}\right)^{2}&=&\frac{8\pi G}{3}\rho
           -\frac{Kc^{2}}{a^{2}},\label{efe1}\\
\frac{\ddot{a}}{a}&=&-\frac{4\pi G}{3}\left(\rho
+3\frac{p}{c^{2}}\right). \label{efe2}
\end{eqnarray}
 Usually, due to the Bianchi identity, one can deduce from
these two equations the equation of energy conservation. In this
new setting, since $\dot c\not=0$, this becomes
\begin{equation}\label{cos}
\dot\rho+3H\left(\rho+\frac{P}{c^2}\right)=\frac{3Kc^2}{4\pi G
a^2} \frac{\dot c}{c}. \label{cons1}
\end{equation}
At this stage, let us say that the set of equations
(\ref{efe1}-\ref{cos}) are a consistent set of equations. But
these equations are just postulated and are not sufficient to
define clearly the theory they derive from. In particular
Eqs.~(\ref{efe1}-\ref{cos}) do not provide an equation for the
evolution of $c$. Let us recall, as Jordan~\cite{jordan} first
pointed out, that it is usually not consistent to allow a constant
to vary in an equation that has been derived from a variational
principle under the hypothesis of this quantity being constant. He
stressed that one needs to go back to the Lagrangian and derive
new equations after having replaced the constant by a dynamical
field.

\subsection{The VSL variational principle}

It was argued~\cite{am,bm,bm2,barrow} that
Eqs.~(\ref{efe1}-\ref{cos}) can be derived from an action
involving a scalar field $\psi \equiv c^4$ given by
\begin{equation}
S=\int\dd x^4\left[\sqrt{-g}\left( \frac{\psi (x^j)}{16\pi
G}(R+2\Lambda )+ \mathcal{L}_M\right) +\mathcal{L}_\psi \right]
\label{act}
\end{equation}
where $\Lambda$ is the cosmological constant. ${\cal L}_M$ is the
matter Lagrangian and $\mathcal{L}_\psi$ controls the dynamics of
$\psi$. The authors of Refs.~\cite{am,bm,bm2,barrow} explain that
the Riemann tensor (and Ricci scalar) is to be computed in one
frame (or at constant $\psi$, as in the usual derivation);
additional terms in $\partial_\mu\psi$ must be present in other
frames. According to this interpretation of the variational
principle, it is argued that the field equations resulting from
(\ref{act}) are
\begin{equation}
G_{\mu\nu }-g_{\mu \nu }\Lambda =\frac{8\pi G}{\psi} T_{\mu \nu }.
\label{feBM}
\end{equation}

However without any additional specification, {\it the action
(\ref{act}) is nothing but a scalar-tensor theory with $F=\psi$}
(cf. the discussion in the previous section) and should be varied {\it
in the usual way}, i.e. using the standard variational methods as
given e.g. in \cite{ryd} (p.84) or \cite{wei} (p.300).  Note that the
resulting Euler-Lagrange equations are the conditions for the action
to be stationary, and the proof that this is so is independent of the
frame chosen for the calculation.  This standard variational principle
leads to the results recalled in Section~\ref{sec_ST}.  Setting
$Z=U=0$ in those equations to match the VSL equations as closely as
possible gives
\begin{equation}
\psi G_{\mu \nu }=8\pi G T_{\mu \nu }+\nabla _{\mu }\partial _{\nu
} \psi-g_{\mu \nu} (g^{\alpha\beta}\nabla_\alpha\nabla_\beta)\psi.
\end{equation}
Furthermore, as can be now easily seen, {\it Eq.~(\ref{feBM}) does not
agree with the Lagrangian~(\ref{act}) varied in the standard way}. To
be specific about this: {\it equations (\ref{feBM}) do not result from
the action given by~(\ref{act}) being stationary under arbitrary
variations when $\psi$ is allowed to be a space-time function}.
Indeed, the cosmological equations resulting from this action (and
with $c_{\rm ST}$ set equal to 1) take the form
\begin{eqnarray}
H^2&=& \frac{8\pi
G}{3\psi}\rho+\frac{1}{6}Z\frac{\dot\psi^2}{\psi}-H\frac{\dot\psi}{\psi}
+\frac{U}{3\psi}-\frac{K}{a^2}\\ \frac{\ddot a}{a}&=& -\frac{4\pi
G}{3\psi}(\rho+3P)-\frac{1}{3}Z\frac{\dot\psi^2}{\psi}
-H\frac{\dot\psi}{2\psi}-\frac{\ddot\psi}{2\psi}+\frac{U}{3\psi}\\
\dot\rho+3H(\rho+3P)&=&0\\
Z\left(\ddot\psi+3H\dot\psi\right)&=&3\left(\dot
H+2H^2+\frac{K}{a^2}\right)-Z'\frac{\dot\psi^2}{2}-U',
\end{eqnarray}
which differ from Eqs.~(\ref{efe1}-\ref{cos}).

Thus the claimed variational principle used in the VSL 
models~\cite{am,bm,bm2,barrow} ,
leading from (\ref{act}) to (\ref{feBM}), is not a standard
variational principle; how this implication can be deduced is
unclear. Let us also add that even in the case in which $c$ is not
varying but $G$ is, the equations (3-6) in which $\dot c=0$ of
Ref.~\cite{barrow} do not reduce to the scalar-tensor equations,
which are the archetype of varying $G$ theories.

\subsection{VSL Field equations}

Let us have an open minded attitude and forget about the VSL
action (\ref{act}), which is useless since it implies a
non-standard and non-defined variational principle. Let us assume
rather that one can postulate the VSL-Einstein field equation
(\ref{feBM}), even if we know this is a dangerous approach, as
explained by Jordan~\cite{jordan}. Then the Bianchi identities
imply that the stress-energy tensor satisfies the conservation
equation
\begin{equation}\label{bi}
\nabla_\mu T^{\mu\nu}=-T^{\mu\nu}\nabla_\mu\psi.
\end{equation}

Postulating the equations (\ref{feBM}) and (\ref{bi}) drives a set
of questions and comments:
\begin{enumerate}
\item Can it be demonstrated that these equations
 can be derived from some kind of
Lagrangian varied in the standard way?
\item From the equations
(\ref{feBM}) and (\ref{bi}), one cannot derive a propagation
equation for $\psi$. It is thus impossible to determine the
degrees of freedom of the theory just on the basis of this set of
equations (is $\psi$ a spin-0 field or just a function?). This
step requires in general writing a Lagrangian (recall the
discussion between Jordan and Einstein frames of
Section~\ref{sec_ST}). In particular, without a Lagrangian, it
seems impossible to decide whether the theory is well defined
(e.g. has no negative energy etc.).
\item How are we sure that $T^{\mu\nu}$ in Eq. (\ref{feBM}) is in
fact the stress energy tensor? Let us recall that in general, if
one has a matter action of the form
\begin{equation}
S_m=\int\sqrt{-g}L(\phi,\partial_\mu\phi,g_{\mu\nu})\dd^4{\bf x}
\end{equation}
that is a scalar, then invariance under any reparameterization
$\xi^\mu$ implies that if the Euler-Lagrange equations are
satisfied then
\begin{equation} \int\frac{\delta\sqrt{-g}L}{\delta
g_{\mu\nu}}{\cal L}_\xi g_{\mu\nu}\dd^4{\bf x}=0
\end{equation}
where ${\cal L}_\xi g_{\mu\nu}$ is the Lie derivative of the
metric. It follows that, for all $\xi^\mu$,
\begin{equation} \int\xi_\mu\nabla_\rho
T^{\rho\mu}\dd^4{\bf x}=0\quad \hbox{with the definition}\quad
T^{\mu\nu}=\frac{2}{\sqrt{-g}}\frac{\delta\sqrt{-g}L}{\delta
g_{\mu\nu}}
\end{equation}
which yields the stress-tensor conservation equation. Thus, what
we call the stress-energy tensor is closely related with the
invariance under coordinate transformations. This definition
agrees with the weak field limit of what we call energy, etc.

When postulating Eq.~(\ref{feBM}), how can one be sure that the
$T^{\mu\nu}$ that appears there agrees with the standard notion of
stress-energy tensor? In particular, one can redefine the
stress-energy tensor as $T^{\mu\nu}/\psi$ in order to make $\psi$
disappear. This new stress-energy tensor will be conserved as
usual, and it would seem natural to identify it as the physical
stress-energy tensor. What would need to be demonstrated to
justify the proposed alternative identification in the VSL papers,
in some sense equivalent to measuring energy and momentum in
different units than usual, is that $\psi$ and $T^{\mu\nu}$ can be
measured independently, and that the latter agrees with the usual
notion of mass etc.
\end{enumerate}

In conclusion, we see that even if one can always postulate
equations, their interpretation is somewhat easier when one has a
Lagrangian, but a Lagrangian relates as usual to standard physics
only when it is varied in the usual way. Besides, the preceding
remarks seem to show that in fact equations (\ref{feBM}) and
(\ref{bi}) describe nothing more than standard general relativity
in weird (spacetime dependent) units. Another hint that this may
be the case is given by the cosmological equations:
Eq.~(\ref{cos}) is the only one to depend on $\dot c$ and then
only if $K\not=0$. We know that if $K=0$, the Euclidean space is
scale-free, so that a homogeneous change of units will not affect
it, while spherical and hyperbolic spaces have a preferred scale
set by their curvature that is affected by a change of units.

\subsection{Dynamics of $\psi$}

Let us come back to the dynamics of $\psi$. As we said, if it is
not dynamical then it is not a true degree of freedom of the
theory and can be eliminated.

It was proposed in the papers referred to that its dynamics is
driven by the Lagrangian
\begin{equation}\label{lpsi}
{\cal L}_\psi=-\frac{\omega}{16\pi G\psi}\dot\psi^2
\end{equation}
in the preferred rest-frame. Note that it is important that this
is not multiplied by $\sqrt{-g}$ in the action (\ref{act}).

Unfortunately, since (as shown above) the standard variational
principle does not apply and we were not given the proposed new
variational principle, one cannot guess the equation of evolution
for $\psi$ from this Lagrangian so that, at this stage, this new
piece of information is useless.

\subsection{Is it possible to formalize the VSL variational principle?}

As emphasized in the original works on VSL, it must be emphasized
that a spacetime dependent speed of light will imply that local
Lorentz invariance is broken. The action (\ref{act}) seems at
first glance Lorentz invariant, but it was argued in the VSL
papers that it holds only in a favored frame, backing up our claim
that VSL theory uses a non-standard variational principle.

There are other cases in which local Lorentz invariance is broken
that are known in physics, e.g. trans-Planckian physics both for
black-holes and cosmology. In those cases, it is well known that
one can write down a Lagrangian~\cite{llmu} that can be varied in
the standard way. This requires the introduction of an auxiliary
timelike vector field $u^\mu$ (with $g_{\mu\nu}u^\mu u^\nu=-1$)
that specifies the preferred frame and according to which an
observer moving with $u^\mu$ (i.e. $\dd x^\mu/\dd s=u^\mu$) can
speak of space and time.

Proceeding in this way, the Lagrangian for the scalar field $\psi$
corresponding to Eq.~(\ref{lpsi}) will take the form
\begin{equation}
{\cal L}_\psi=-\frac{\omega}{16\pi
G\psi^{3/2}}(u^\mu\partial_\mu\psi)^2
\end{equation}
(assuming that the dot in Eq.~(\ref{lpsi}) referred to a
derivative with respect to the proper time of an observer moving
with $u^\mu$). Then, to express the Einstein Lagrangian in term of
$u^\mu$, one would need to use the Gauss equation and introduce
the metric of the 3-space as defined by $u^\mu$
\begin{equation}
\gamma_{\mu\nu}=g_{\mu\nu}+u_\mu u_\nu.
\end{equation}
We will not go further, since it will imply trying to define for
instance what we mean by speed of light, causality etc. in this
context and many choices are (in principle) possible. In any case,
one would get an action of the type
\begin{eqnarray}
S&=&\int\dd x^4\sqrt{-g} \left(\frac{\psi}{16\pi G}
  (R[\gamma_{\mu\nu},\psi,u^\mu]+2\Lambda)
 +\mathcal{L}_M +\lambda(u^\mu u_\mu+1)+\mathcal{L}_u
\right) \nonumber \\
&-&\int\dd x^4 \left[\frac{\omega} {16\pi
G\psi^{3/2}}(u^\mu\partial_\mu\psi)^2 \right] \label{actalaJP}
\end{eqnarray}
where $\lambda$ is a Lagrange multiplier to ensure the
normalization of $u^\mu$ during the variation and $\mathcal{L}_u$
is the Lagrangian describing the dynamics of $u^\mu$. Such a
Lagrangian is covariant but involves a preferred frame so that it
can be a starting point to formulate a varying speed of light
theory. Clearly, it is much more complicated and involved than the
naive version (\ref{act}) and will certainly not lead to field
equations of the simple form (\ref{feBM}).

\subsection{Conclusions}

The variational principle used in the VSL theories in\cite{am,bm,bm2,barrow} is 
clearly
non-standard. The field equations postulated seem in fact not to
describe anything but general relativity itself (written with
strange units), while the action proposed (if varied in the usual
way) would correspond to a usual scalar-tensor theory. We stress
that it is possible to write variational theories in which local
Lorentz invariance is broken, but than this is much more involved
that what is presented in Refs.~\cite{am,bm,bm2,barrow}.

Some other comments are needed. It seems that before calling a
theory `varying speed of light', it would have been necessary to
make explicit which speed of light was meant, and indeed it should
have been related to both major meanings ($c_{_{\rm EM}}$ and
$c_{_{\rm ST}}$) via the relevant equations. This was not done;
for instance, no study of electrodynamics was presented in these
papers. Most of the advertisement for the theory relies on the
cosmological equations. It is indeed odd to discuss the cosmology
of a claimed varying-$c$ theory that has not been defined properly
in relation to the usual meanings of the speed of light, and
related to standard physics of measurement and electromagnetism.
Thus, it cannot be taken seriously, according to usual theoretical
standards, as a proposal for the speed of light to vary. According
to what we explained before, it also seems that the resolution of
the e.g. horizon problem arises from the fact that one compared
clocks with different dials, rather than providing either a
mechanism for the speed $c_{_{\rm EM}}$ of electromagnetic waves
to vary or for the limiting causal speed $c_{_{\rm ST}}$ to vary
(cf. the discussion of Moffat's bimetric theory below; see also 
Ref.~\cite{coule} for earlier arguments).

As to Moffat's original theory \cite{mof1}, it has a much more
sophisticated variational principle than the papers discussed above,
but also gives us no reason as to either why the physical speed of
light (as determined by Maxwell's equations) should vary, nor any
reason as to why an arbitrarily introduced time coordinate time $t$
should be regarded as having physical meaning, rather than proper time
$\tau = \int \sqrt{-\dd s^2}$ determined by the metric tensor. Indeed
no physical characterisation is given whereby the coordinate time $t$
in Moffat's equations (26) and (46) can be determined by some
measurement process.

A last comment. If the speed of light is dynamical and is
self-consistently replaced by a dynamical field (or in fact many,
relaxing the standard concordance of all the facets of $c$), then
these fields will couple to ordinary matter. To discuss the effect
of the new forces that will appear and in particular the possible
violation of the universality of free fall, one requires to study
also the implications of the theory in the Solar System. It was
assumed that the preferred frame is the cosmological frame in
which the Friedmann equations hold (indeed expansion breaks
Lorentz invariance); this does not easily allow one to perform
such a Solar System study.

\section{Other implementations: Bimetric theories}\label{sec_moffat}

\subsection{Bekenstein type theories}

Among the other implementations of a varying fine structure
constant is a theory of varying electric charge formulated about
20 years ago by Bekenstein~\cite{be}. In that case, the speed of
light and Planck constant are supposed to be fixed so that one
gets an unambiguous varying fine structure constant theory in
which a scalar field couples to the electromagnetic Lagrangian.
This was rephrased in Ref.~\cite{bssm} after a field redefinition
(hence leading to the same theory). In Ref.~\cite{jm}, a covariant
form of the VSL theory was proposed, based on the introduction of
a field $\chi$ as $c=c_0\exp\chi$ and assuming that the full
matter Lagrangian does not contain $\chi$, leading to an action of
the type
\begin{equation}
S=\int\dd^4{\bf x}\sqrt{-g}\left(\frac{R}{16\pi
G}-\frac{\omega}{2}\partial_\mu\chi\partial^\mu\chi+{\cal
L}_{M}\hbox{e}^{b\chi}\right).
\end{equation}
Is this theory supposed to be a better version of the one discussed
before, or a new theory? What facet of the speed of light is supposed
to be varying? For instance, the latter Lagrangian is Lorentz
invariant, contrary to all the discussions about the breakdown of this
invariance in VSL~\cite{am,bm,bm2,barrow}. As correctly studied in
Ref.~\cite{mbs}, this theory has different observational implications
than the Bekenstein theory~\cite{be} or its rephrasing~\cite{bssm}. It
seems however exaggerated to conclude that either $c$ or $e$ is
varying, at least as long as one has not defined properly what is
meant by $c$. What is being examined in these papers, as far as we can
judge from the Lagrangians, are different scalar-tensor theories with
different kinds of coupling between standard matter and the scalar
field. Although this is called a VSL theory, it does not seem to 
provide any way to solve the horizon problem in cosmology.

\subsection{Bimetric theories}

An alternative proposal is bimetric theories.  Here we consider the
geometric aspects of Moffat's varying-$c$ theory as set out in
\cite{moff1}.  In this paper a `bimetric form' is proposed for the
cosmological metric $g_{\mu \nu }$ as follows [equations (18)-(20)]:
\begin{equation}
g_{\mu \nu }=g_{0\,\mu \nu }+g_{m\,\mu \nu }  \label{bimetric}
\end{equation}
where
\begin{equation}
g_{0\,\mu \nu }\dd x^{\mu }\dd x^{\nu }=\dd t^{2}c_{0}^{2}\theta
(t_{c}-t)-R^{2}(t) \left[ \frac{\dd
r^{2}}{1+kr^{2}}+r^{2}\dd\Omega^{2}\right]  \label{metric1}
\end{equation}
\begin{equation}
g_{m\,\mu \nu }\dd x^{\mu }\dd x^{\nu }=\dd t^{2}c_{m}^{2}\theta
(t-t_{c})-R^{2}(t) \left[ \frac{\dd
r^{2}}{1+kr^{2}}+r^{2}\dd\Omega^{2}\right]  \label{metric2}
\end{equation}
Here $\dd\Omega^{2}=\dd\vartheta^{2}+\sin^{2}\vartheta
\dd\varphi^{2}$ and $\theta (t)$ is the Heaviside stepfunction:
$\theta (t)=1$ for $t>0$ and $\theta (t)=0$ for $t<0.$ It is
suggested that this leads to a phase transition in the speed of
light at a time $t=t_{c}$:
\[
c(t)=c_{0}\theta (t_{c}-t)+c_{m}\theta (t-t_{c}).
\]
Now the metric (\ref{metric1}) is degenerate for $t>t_{c}:$ it has
no time dimension then, and similarly, the metric (\ref{metric2})
is degenerate for $t<t_{c}$ The final metric form (\ref{bimetric})
is a single metric with an apparent jump in the speed of light at
time $t=t_{c}.$ However this change does not represent a change in
the physical speed of light. In the metric (\ref{metric1}), define
proper time $\tau $ by $2\dd\tau ^{2}=c_{0}^{2}\dd t^{2}\ $ and
similarly by $2\dd\tau ^{2}=c_{m}^{2}\dd t^{2}$ in the metric
(\ref{metric2}). Then on using the proper time $\tau \,$ in each
metric instead of the arbitrary coordinate time $t,$ the metric
form (\ref{bimetric}) will be
\begin{equation}
g_{\,\mu \nu }\dd x^{\mu }\dd x^{\nu }=2\dd\tau ^{2}-2R^{2}(\tau
)\left[ \frac{\dd r^{2} }{1+kr^{2}}+r^{2}\dd\Omega ^{2}\right]
\label{cont}
\end{equation}
for all values of $\tau .$ There is no jump in the final metric
form (\ref {bimetric}), and indeed $\tau \,$ is just the standard
physically measurable proper time in this continuous metric, up to
a factor $2$ (needed because the spatial part of the metric will
occur there with a factor 2). As usual, the physical speed of
light will take the constant value $1$ at all times in these
physical coordinates. We are given no physical reason to prefer 
the arbitrary coordinate time $t$ in these equations over proper 
time $\tau$ as determined in the standard way by the metric tensor.

Indeed, this is not really a bimetric form at all, i.e. there are not two
separate metrics in use at each time with separate physical
interpretations, rather in (\ref{metric1}-\ref{metric2}) we have a
single metric written as the sum of two parts that vary in the two
coordinate patches. A single coordinate patch can be found that
eliminates the coordinate singularity and apparent change in the
speed of light at the time $t=t_{c}$.

By contrast, the theories of Bekenstein~\cite{be1}, Clayton and
Moffat~\cite{claymof}, and Bassett {\em et al.}~\cite{bass} are
genuine bimetric theories: they have one metric governing
gravitational phenomena and another governing matter~\cite{claymof} or
explicitly electromagnetic effects~\cite{bass}. Consequently these are
much more complex than standard general relativity, needing field
equations for both metrics (or equivalently, for one metric and for a
vector or scalar that determines the difference between the metrics),
hence raising many issues about gravitational lensing and the
equivalence principle (see e.g. Ref.~\cite{will} for
discussions). However this does provide a genuine physical basis for
discussing the differences between the various facets of $c$ discussed
in this paper.

\section{Conclusions}\label{sec_concl}

To conclude, we have tried to recall that the nature of the speed
of light is complex and has many facets. These different facets
have to be distinguished if one wants to construct a theory in
which it is supposed to vary. In particular if it is the
electromagnetic speed that is supposed to vary, we should be shown
how Maxwell's equations are to be changed; if it is the causal
speed that is to vary we should be shown how the spacetime metric
tensor structure and interpretation is altered.

As we emphasized, letting a constant vary implies replacing it by
a dynamical field consistently. Letting it vary in equations
derived under the assumption it is a constant leads to incorrect
results, as the example of a scalar tensor theory clearly shows.
One needs to go back to a Lagrangian that allows one to determine
the degrees of freedom of the theory and to check if it is well
defined.

Concerning the VSL theories in~\cite{am,bm,bm2,barrow}, 
let us recall that the variational
principle used is clearly non-standard and that the field
equations that are postulated seem in fact not to describe
anything else but general relativity itself in unusual units. The
emphasis must be put on what can be measured, and in that respect
considering only the variation of dimensionless quantities makes
sense.

The possibility that the fundamental constants may vary during the
evolution of the universe offers an exceptional window onto higher
dimensional theories and is probably linked with the nature of the
dark energy that makes the universe accelerate today. Thus the
topic is worth investigating. Physics is however about precise
words and clear concepts, and it would be a pity to let the
present confusion over the nature of varying-$c$ theories spread.
It is also a pity that advocates of VSL theories choose to use 
propaganda methods, using phrases such as ``religious fervour''  and 
``risible'' \cite{propaganda},  rather than addressing the central 
scientific issues that arise, such as how distances may be 
measured with high accuracy independently of the speed of light.
We hope that the basics recalled in this text will help in this
direction.

\section*{Acknowledgements}

We thank Gilles Esposito-Far\`ese for many discussion on Lagrangians
and their use in physics, J\'er\^ome Martin, Filippo Vernizzi, Gilles
Cohen-Tannoudji and Ren\'e Cuillierier for discussions on the
constants of Nature. We also thank Joao Magueijo, John Barrow, John
Moffat and Michael Clayton for commenting on their work and Christian
Armendariz-Picon, David Coule, Vittorio Canuto and Thomas Dent. JPU
thanks the University of Buenos Aires, and particularly Diego Harari,
for hospitality during the latest stages of this work. GE thanks the
NRF and University of Cape Town for support.


\end{document}